\documentstyle[aps,twocolumn,prb,epsf]{revtex} 

\begin{document}

\draft
\twocolumn[    
\hsize\textwidth\columnwidth\hsize\csname @twocolumnfalse\endcsname    

\title{Optimizing thermal transport in the Falicov-Kimball model: binary-alloy 
picture
}
\author{J. K. Freericks$^{1}$ and V. Zlati\'c$^2$}
\address{$^1$Department of Physics, Georgetown University, 
  Washington, DC 20057-0995, U.S.A.\\
$^2$Institute of Physics, Zagreb, Croatia}
\date{\today}
\maketitle

\widetext
\begin{abstract}
We analyze the thermal transport properties of the Falicov-Kimball model
concentrating on locating regions of parameter space where the thermoelectric
figure-of-merit $ZT$ is large.  We focus on high temperature for power
generation applications and low temperature for cooling applications.  
We constrain the
static particles (ions) to have a fixed concentration, and vary the 
conduction electron concentration as in the binary-alloy picture of the
Falicov-Kimball model.  We find a large region of parameter space with
$ZT>1$ at high temperature and we find a small region of parameter space
with $ZT>1$ at low temperature
for correlated systems, but we believe inclusion of the lattice thermal
conductivity will greatly reduce the low-temperature
figure-of-merit.
\end{abstract}

]      

\narrowtext
\section{Introduction}

There has been a recent resurgence of interest in solid-state devices for
thermoelectric applications\cite{mahan_phystoday}
(power generation or cooling).  The recent focus
has been on investigating strongly correlated materials, which may prove to
have better performance than the current generation of semiconductor-based
devices.  The two main areas of application for thermoelectrics are in
power generation from the Peltier effect,\cite{peltier}
where heat energy is converted
into electricity; and in thermoelectric cooling, where an electrical
current is driven through a device to force heat to move from the cold to hot
end.  Power generation applications typically operate at temperatures higher 
than 1000~K, with the heat source being a radioactive material (for applications
in the space industry) or a combustion source.  Thermoelectric coolers
usually operate around room temperature, and the semiconductor-based devices
do not function below about 200~K.  Currently, thermoelectric devices fit niche
markets, where reliability, size,  or weight are more important than efficiency.
The coolers usually operate with relatively low heat loads because of
their poor efficiency.

The efficiency of a thermoelectric device is a function of the dimensionless
product of a materials parameter denoted $Z$ with the average temperature $T$ 
(between the hot and cold heat sources of the device) and
is called $ZT$.  It satisfies
\begin{equation}
ZT=\frac{T\sigma_{dc}S^2}{\kappa_e+\kappa_l},
\label{eq: ztdef}
\end{equation}
and the term $\sigma_{dc}S^2$ in the numerator is often called the power factor.
Here we have $\sigma_{dc}$ the dc electrical conductivity, $S$ is the 
Seebeck coefficient,\cite{seebeck} $T$ is the temperature, $\kappa_e$ is the
electronic contribution to the thermal conductivity, and $\kappa_l$
is the lattice contribution to the thermal conductivity (we are assuming the
electron-phonon interaction is small enough that these two effects can be
decoupled).  It is commonly stated that $ZT>1$ is needed for operation of
thermoelectric devices, but this is not necessarily true.\cite{rowe_bhandari}
For example,
if we consider a thermoelectric cooler operating at 300~K and with 50~K of
cooling, then one can operate such a device for $ZT>0.7$, but one would need
$ZT\approx 4$ to achieve the coefficient of performance of a conventional
compressor-based refrigerator (which lies in the $1.2-1.4$ range).
Nevertheless, most commercial thermoelectric
devices have $ZT$ near 1 because no materials have been discovered
with much larger values.  Of course there is significant interest
in increasing $ZT$ by a factor of 4 at room
temperature (to be competitive with conventional coolant-based
technology), or to above 1 at low temperature to allow for 
new applications such as solid-state coolers for superconducting electronics
or infrared detectors.

Although there are no fundamental thermodynamic limitations\cite{mahan_review}
to the size of $ZT$, it has proved to be quite difficult to find materials with
$ZT>1$ over a wide temperature range, and to find much larger
values of $ZT$ (say $ZT>3$).  Recently, Rontani and Sham\cite{rontani_sham}
proposed that heterostructures
of correlated semiconductors and metals could have dramatically large values
of $ZT$ at low temperature.  Their idea was that if one tuned the large
electronic density of states (DOS) of the $f$-electrons to lie close
to the fermi level, then one could produce huge values of $ZT$ (earlier
work proposed similar ideas as well\cite{mao_bedell}).  Mahan
and Sofo\cite{mahan_sofo} also argued in 
the same vein for optimization in bulk materials in 1996.  But so far,
no one has been able to demonstrate that such large values of $ZT$ are
possible in a true many-body system (and are not the artifact of some
approximations employed in the analysis).  We examine this problem in
detail for the Falicov-Kimball model here.  By working in the limit where
the ion concentration is fixed and nonzero at $T=0$, we have adjusted the
renormalized energy level of the ion to lie at the electronic chemical
potential, which has the potential for producing large thermoelectric
responses.

Mahan and Sofo\cite{mahan_sofo}
also proved that the figure-of-merit always satisfies an inequality
\begin{equation}
ZT<\frac{\kappa_e+T\sigma_{dc}S^2}{\kappa_l}
\label{eq: ztineq}
\end{equation}
regardless of the strength of the many-body interactions.  This has
important implications for theorists, because in purely electronic models,
such as the one we investigate here, $\kappa_l=0$, so there is no
{\it a priori} limitation on the magnitude of $ZT$.  But it also
presents some problems for low-$T$ calculations, since the electronic
contribution to the thermal conductivity is usually much smaller than the 
lattice contribution at low temperature (especially for insulators), and hence 
{\it purely electronic estimates
of $ZT$ can be greatly enhanced when the lattice effects are ignored.} This
becomes less of an issue at high temperature, where the electronic contribution
to the thermal conductivity can dominate.

Another interesting feature that plagues the low-$T$ regime is the fact that
in most systems $S\rightarrow 0$ as $T\rightarrow 0$.  Since the ratio
of the conductivities often satisfies the Wiedemann-Franz law
\begin{equation}
\frac{\kappa_e}{\sigma_{dc}}=\left (\frac{k_B}{e}\right )^2 {\mathcal L}T
\label{eq: lorenz}
\end{equation}
with ${\mathcal L}$ the Lorenz number (equal to $\pi^2/3$ in a fermi
liquid and 3 in an intrinsic semiconductor), we have $ZT\rightarrow 0$
if $S\rightarrow 0$ at low temperature.  Similarly, if $S(T)$ suffers
a sign change at  any $T$, then $ZT$ will be quite low in the vicinity of
the sign change.

The Falicov-Kimball model\cite{falicov_kimball}
appears to be able to describe an increasing number of
materials and systems.  One example, that fits within the binary-alloy picture,
is tantalum defficient tantalum nitride\cite{newman}
Ta$_x$N.  This material is metallic
when $x=1$ but becomes a fairly large-gap insulator (about 1.5~ev) when
$x=0.6$.  If we let the $A$ ion denote a unit cell with a Ta atom, and a $B$ ion
denote a unit cell with no Ta atom, then $U$ is the difference in site energies
for the two configurations.  The total conduction electron concentration also
depends on the Ta vacancies, as each vacancy can bind 5 electrons. It is easy
to model the metal-insulator transition at $x=0.6$ by properly varying the 
electron concentration with the concentration of Ta vacancies.

In Section II we develop the formalism for deriving the dc conductivity, the
thermopower and the thermal conductivity.   We use this to determine both
the Lorenz number and the figure-of-merit.
In Section III we provide numerical results for the thermal
transport illustrating regimes where $ZT$ can become large, and describing
the physical principles that drive such enhancements.  In addition, we
describe in detail the situation behind a large figure-of-merit at low-$T$
and whether achieving such a goal is feasible.
Conclusions are presented in Section IV.

\section{Formalism for the thermal transport}

The Hamiltonian we study is the spinless Falicov-Kimball 
model\cite{falicov_kimball} with a canonical-binary-alloy picture
\begin{equation}
H=-\frac{t^*}{2\sqrt{d}}\sum_{\langle i,j\rangle}c^\dagger_{i}
c_{j}+U\sum_{i}w_ic^{\dagger}_{i}c_{i},
\label{eq: ham}
\end{equation}
where $c^{\dagger}_{i}$ ($c_{i})$ is the electron creation 
(annihilation)
operator for a spinless electron at site $i$ (spin can be included trivially
if desired by doubling all of the $L_{ij}$ defined below), 
$w_i$ is a variable that equals
zero or one and corresponds to the presence of an $A$ ion ($w_i=1$) or the
presence of a $B$ ion ($w_i=0$) at site-$i$, and $U$ is
the interaction strength (difference in the site energy between the 
$A$ and $B$ ions).  The hopping integral is scaled with the spatial
dimension $d$ so as to have a finite result in the limit\cite{metzner_vollhardt}
$d\rightarrow \infty$; we measure all energies in units of $t^*=1$. We work
on a hypercubic lattice where the noninteracting density of states is
a Gaussian $\rho(\epsilon)=\exp(-\epsilon^2)/\sqrt{\pi}t^*\Omega_{uc}$
(with $\Omega_{uc}$ the volume of the unit cell). A chemical 
potential $\mu$ is employed to adjust the conduction electron filling $\rho_e$.

The Falicov-Kimball model can be solved exactly by employing dynamical
mean field theory\cite{brandt_mielsch,freericks_fk}. A review that describes
how to solve for the Green's function using the equation of motion technique
has recently appeared.\cite{freericks_review}
Because the self energy $\Sigma(z)$ has no momentum 
dependence, the local Green's function satisfies
\begin{equation}
G(z)=\int d\epsilon \rho(\epsilon)\frac{1}{z+\mu-\Sigma(z)-\epsilon},
\label{eq: g_loc}
\end{equation}
with $z$ a complex variable in the complex plane.  
The self energy, local Green's 
function, and effective medium $G_0$ are related to each other by
\begin{equation}
G_0^{-1}(z)-G^{-1}(z)=\Sigma(z),
\label{eq: g_0}
\end{equation}
and the Green's function also satisfies
\begin{equation}
G(z)=(1-w_1)G_0(z)+w_1\frac{1}{G_0^{-1}(z)-U}.
\label{eq: g_atomic}
\end{equation}
Here $w_1$ is the average concentration of the A ions (which is an input
parameter).  The algorithm for determining the Green's function begins with
the self energy set equal to zero.  Then Eq.~(\ref{eq: g_loc}) is used
to find the local Green's function.  The effective medium is found from
Eq.~(\ref{eq: g_0}).
The new local Green's function is then found from Eq.~(\ref{eq: g_atomic})
and the new self energy from Eq.~(\ref{eq: g_0}).  This algorithm is 
repeated until it converges.

When these equations are solved we find a number of interesting results
for the single-particle properties.  First, both the interacting DOS
[$\rho_{int}(\omega)=-{\rm Im}\{G(\omega+i\delta)\}/\pi$]  and the 
self energy on the real axis are independent of 
temperature\cite{vandongen_leinung}
when $w_1$ and $\mu$ are fixed (all of the temperature dependence for fixed 
$\rho_e$ arises from the temperature dependence of $\mu$, which shifts the
zero-frequency point of the DOS). Second, we find that the self energy does not
display fermi-liquid properties unless $U=0$, $w_1=0$ or $w_1=1$ (which are
all noninteracting cases).  In particular, we do find (for small enough $U$)
that the imaginary
part of the self energy is quadratic in $\omega$, but the curvature has
the wrong sign, and the zero frequency value of the imaginary part of the
self energy does not go to zero as 
$T\rightarrow 0$ (in fact, it remains fixed for all $T$). Third, the
real part of the self energy is linear (for small enough $U$), but the slope
has the opposite sign of what is seen in a fermi liquid.  Finally, we 
see that if $U$ is large enough, the DOS develops a gap, and if the electronic
chemical potential lies in the gap, then the self energy has quite anomalous 
behavior (including developing a pole).

Transport properties are calculated within a Kubo-Greenwood 
formalism\cite{kubo}.  This
relates the transport coefficients to correlation functions of the 
corresponding transport current operators (these are equal to the bare
bubbles because there are no vertex corrections in the large-dimensional
limit~\cite{khurana}).  We need two
current operators here---the particle current\cite{mahan_text}
\begin{equation}
{\bf j}=\sum_{q}{\bf v}_q c^\dagger_{q}c_{q},
\label{eq: j_particle}
\end{equation}
(where the velocity operator is ${\bf v}_q=\nabla_q\epsilon(q)$, the 
bandstructure is $\epsilon(q)$, and the
Fourier transform of the creation operator is $c^\dagger_q=\sum_j
\exp[iq\cdot {\bf R}_j]c^\dagger_j$/N) and the heat 
current\cite{jonson_mahan,mahan_text}
\begin{eqnarray}
{\bf j}_Q&=&\sum_{q}[\epsilon(q)-\mu]{\bf v}_q c^\dagger_{q}c_{q}
\cr
&+&\frac{U}{2}\sum_{qq^\prime}W(q-q^\prime)[{\bf v}_q+{\bf v}_{q^\prime}]
c^\dagger_{q}c_{q^\prime},
\label{eq: j_heat}
\end{eqnarray}
[where  $W(q)=\sum_j\exp(-iq\cdot{\bf R}_j) w_j/N$]. 

The dc conductivity, thermopower and electronic thermal conductivity
can all be determined from relevant correlation functions of the
current operators.  We define three transport coefficients $L_{11}$, $L_{12}=
L_{21}$, and $L_{22}$. Then
\begin{equation}
\sigma_{dc}=e^2L_{11},
\label{eq: conductivity}
\end{equation}
\begin{equation}
S=\frac{k_B}{|e|T}\frac{L_{12}}{L_{11}},
\label{eq: thermopower}
\end{equation}
and
\begin{equation}
\kappa_e=\frac{k_B^2}{T}\left [ L_{22}-\frac{L_{12}L_{21}}{L_{11}}\right ].
\label{eq: thermalconductivity}
\end{equation}
One finds that the electric and thermal conductivities are always positive,
but the thermopower can have either sign---a positive thermopower corresponds
to electron-like transport and a negative thermopower to hole-like transport
(we use the sign convention of Ashcroft and Mermin~\cite{ashcroft_mermin}).
The transport coefficients are found from the analytic continuation of
the relevant ``polarization operators'' at zero frequency
\begin{eqnarray}
L_{11}&=&\lim_{\nu\rightarrow 0}{\rm Re}\frac{i}{\nu}\bar L_{11}(\nu),\cr
\bar L_{11}(i\nu_l)&=&\int_0^{\beta}d\tau e^{i\nu_l\tau}
\langle T_\tau j_{\alpha}^\dagger(\tau)j_{\beta}(0)\rangle,
\label{eq: l11}
\end{eqnarray}
where $\nu_l=2\pi Tl$ is the Bosonic Matsubara frequency
($\beta=1/T$), the $\tau$-dependence
of the operator is with respect to the full Hamiltonian in Eq.~(\ref{eq: ham}),
and we must analytically continue $\bar L_{11}(i\nu_l)$ to the real
axis $\bar L_{11}(\nu)$ before taking the limit $\nu\rightarrow 0$.  Similar
definitions hold for the other transport coefficients:
\begin{eqnarray}
L_{12}&=&\lim_{\nu\rightarrow 0}{\rm Re}\frac{i}{\nu}\bar L_{12}(\nu),\cr
\bar L_{12}(i\nu_l)&=&\int_0^{\beta}d\tau e^{i\nu_l\tau}
\langle T_\tau j_{\alpha}^\dagger(\tau)j_{Q\beta}(0)\rangle,
\label{eq: l12}
\end{eqnarray}
and
\begin{eqnarray}
L_{22}&=&\lim_{\nu\rightarrow 0}{\rm Re}\frac{i}{\nu}\bar L_{22}(\nu),\cr
\bar L_{22}(i\nu_l)&=&\int_0^{\beta}d\tau e^{i\nu_l\tau}
\langle T_\tau j_{Q\alpha}^\dagger(\tau)j_{Q\beta}(0)\rangle.
\label{eq: l22}
\end{eqnarray}
In all of these equations, the subscripts $\alpha$ and $\beta$ denote the
respective spatial index of the current vectors. Note that our definitions of
the $L_{ij}$ coefficients, while standard,
have one less factor of $T$ than seen in some other papers.

The analytic continuation is straightforward, and produces the Mott 
form\cite{mott} 
for the transport coefficients (which is usually called the Jonson-Mahan
theorem,\cite{jonson_mahan} and was explicitly evaluated for the 
Falicov-Kimball model\cite{freericks_zlatic}):
\begin{equation}
L_{ij}=\frac{\sigma_0}{e^2}\int_{-\infty}^{\infty}d\omega
\left ( -\frac{df(\omega)}{d\omega} \right ) \tau(\omega)\omega^{i+j-2},
\label{eq: lij_final}
\end{equation}
with the relaxation time $\tau(\omega)$ defined by
\begin{equation}
\tau(\omega)=\frac{{\rm Im}G(\omega)}{{\rm Im}\Sigma(\omega)}+2-2
{\rm Re}\{ [\omega+\mu-\Sigma(\omega)]G(\omega)\}
\label{eq: tau}
\end{equation}      
$f(\omega)=1/[1+\exp(\beta\omega)]$ and $\sigma_0=e^2\pi^2/hda^{d-2}$
on a hypercubic lattice in $d$-dimensions.
Note that even though we represented the above form by an effective relaxation
time, the above expression for the transport coefficients is {\it exact}.

Once the transport coefficients are known, we can determine the electrical
and thermal conductivities and the thermopower.  In addition, we find the 
figure-of-merit satisfies
\begin{equation}
ZT=\frac{L_{12}^2}{L_{11}L_{22}-L_{12}^2}
\label{eq: fom_final}
\end{equation}
when we neglect $\kappa_l$
and the Lorenz number becomes
\begin{equation}
{\mathcal L}\left (\frac{k_B}{e}\right )^2=\frac{\kappa_e}{\sigma_{dc}T}=
\left (\frac{k_B}{e}\right )^2\frac{L_{11}L_{22}-L_{12}^2}{L_{11}^2T^2}.
\label{eq: lorenz_final}
\end{equation}

Note that because $-d f(\omega)/d\omega$ is an even function of $\omega$,
only the even part of $\tau(\omega)$, $\tau_e(\omega)=[\tau(\omega)+
\tau(-\omega)]/2$, contributes to the $L_{11}$ and $L_{22}$ coefficients,
and only the odd part $\tau_o(\omega)=[\tau(\omega)-\tau(-\omega)]/2$ 
contributes to $L_{12}$.  Furthermore, one expects on general physical
grounds that $\tau(\omega)\ge 0$.  Indeed, this can be easily shown to be
true by rearranging Eq.~(\ref{eq: tau}) into
\begin{equation}
\tau(\omega)=\int d\epsilon \rho(\epsilon)\frac{2[{\rm Im}\Sigma(\omega)]^2}
{\{[\omega+\mu-{\rm Re}\Sigma(\omega)-\epsilon]^2+
[{\rm Im}\Sigma(\omega)]^2\}^2}
\label{eq: tau_pos}
\end{equation}
which is manifestly nonnegative. 
In addition, we see that $\tau(\omega)$ vanishes whenever
${\rm Im}\Sigma(\omega)=0$.  In the case of particle-hole symmetry,
when $\rho_e=w_1=0.5$, it is easy to show that $\tau_o(\omega)=0$, and
both the thermopower and $ZT$ must vanish.

We can employ the nonnegativity of $\tau(\omega)$ to show the 
Mahan-Sofo\cite{mahan_sofo}
bound for $ZT$.  The starting point is the fact that
\begin{equation}
\frac{\sigma_0}{e^2}\int_{-\infty}^{\infty}d\omega
\left ( -\frac{df(\omega)}{d\omega} \right ) \tau(\omega)(\omega+a)^2\ge 0
\label{eq: l_ineq}
\end{equation}
which holds since all terms in the integrand are positive.
Choosing $a=L_{12}/L_{11}$ then yields $L_{11}L_{22}-L_{12}^2\ge 0$
(equality only occurs if the integrand is a delta function, which has
no variance).  Now the full expression for $ZT$ can be written as
\begin{equation}
ZT= \frac{L_{12}^2/L_{11}L_{22}}{1-L_{12}^2/(L_{11}L_{22})+
\kappa_lT/(k_B^2L_{22})}=\frac{\xi}{1-\xi+A}
\label{eq: zt_ms}
\end{equation}
by dividing the numerator and denominator in Eq.~(\ref{eq: ztdef}) by 
$k_B^2L_{22}/T$.  Here we used the notation $\xi=L_{12}^2/L_{11}L_{22}$ and
$A=\kappa_lT/(k_B^2L_{22})$.  Using the fact that $\xi\le 1$ then tells us
that $\xi A\le A\le 1-\xi+A$ since $1-\xi\ge 0$.  Dividing both sides
of the inequality by $A(1-\xi+A)$ then yields
\begin{equation}
ZT=\frac{\xi}{1-\xi+A}\le \frac{1}{A}=\frac{\kappa_e+T\sigma_{dc}S^2}{\kappa_l}
\label{eq: zt_ineq_final}
\end{equation}
which is the Mahan-Sofo result.  The inequality in Eq.~(\ref{eq: zt_ineq_final})
holds for any system that satisfies the Jonson-Mahan theorem, regardless of 
the many-body interactions present,
as long as there is no electron-phonon coupling, which precludes the
separation of the thermal conductivity into electronic and lattice pieces.

The integrals for the transport coefficients all have a derivative of the
fermi function in them.  This derivative becomes strongly peaked around
$\omega=0$ with a width on the order of $T$ as $T\rightarrow 0$.  In metals,
we typically find that the relaxation time can be written in the form
$\tau(\omega)=\tau_0+w\tau^\prime+O(\omega^2)$ for small $\omega$.  If we
include only the first two terms in the expansion for the relaxation time,
we find that
\begin{eqnarray}
\sigma_{dc}&=&\sigma_0\tau_0\cr
S&=&\frac{k_B}{|e|}\frac{\pi^2\tau^\prime T}{3}\cr
\kappa_e&=&\frac{k_B^2}{e^2} \frac{\pi^2 \sigma_0\tau_0 T}{3}
\label{eq: transport_metal}
\end{eqnarray}
to lowest-order in $T$.  We can read off that the Lorenz number ${\mathcal L}$
is equal to $\pi^2/3$ and $ZT\rightarrow \pi^2\tau^{\prime 2} T^2
/3$ (when we
neglect $\kappa_l$).  Hence,
$ZT$ will be small in metals at low temperature because Wiedemann-Franz
law fixes the ratio of conductivities, and the thermopower vanishes as
$T\rightarrow 0$.  If the Wiedemann-Franz rule holds at higher temperature
as well, then one needs to find a thermopower larger than $\pi k_B/\sqrt{3}|e|$
which is equal to 156~$\mu$V/K in order to have $ZT>1$ in a metal.  There
are no known metals that have thermopowers larger than 125~$\mu$V/K and most
metals are one to two orders of magnitude smaller,\cite{mahan_review} so the
only way to find metals that are useful for thermoelectric power generation
or cooling is if the Wiedemann-Franz law does not hold.

The situation in an insulator is quite different.  If we assume that 
$\tau(\omega)=\tau_0\rho_{int}(\omega)$ with $\tau_0$ a constant, and if we 
choose a generic interacting density of states that increases like a power law,
\begin{eqnarray}
\rho_{int}(\omega)&=&\theta(\omega-E_g/2)C(\omega-E_g/2)^\alpha\cr
&+&\theta(-\omega-E_g/2)
C^\prime (-\omega-E_g/2)^{\alpha^\prime}
\label{eq: dos_ins}
\end{eqnarray}
[with $\theta(x)$ the unit step function and $E_g$ the insulating gap],
then the behavior of the transport 
coefficients differs from Eq.~(\ref{eq: transport_metal}).  The starting point
is to note that the number of holes excited in the lower band is equal to
the number of electrons in the upper band, and at low temperature the fermi
factors can be replaced by Boltzmann factors, resulting in 
\begin{equation}
\int_{-\infty}^0d\omega\rho_{int}(\omega)e^{\beta(\omega-\Delta\mu)}=
\int_0^{\infty}d\omega\rho_{int}(\omega)e^{-\beta(\omega-\Delta\mu)}
\label{eq: mu_equality}
\end{equation}
with the chemical potential written as $\mu=\mu_0+\Delta\mu(T)$.
Solving for $\Delta \mu(T)$ yields
\begin{eqnarray}
\Delta\mu(T)&\approx&\frac{T}{2}\ln \left [ \int_{-\infty}^0 d\omega
\rho_{int}(\omega)
\exp(\beta\omega)\right ]\cr
&-&\frac{T}{2}\ln \left [ \int_{0}^{\infty} d\omega\rho_{int}(\omega)
\exp(-\beta\omega)\right ]
\label{eq: deltamu_def}
\end{eqnarray}
with $\mu_0$ the chemical potential at $T=0$, which lies in the middle of
the insulating gap (and is chosen to be the origin here).  Using the form for
the DOS in Eq.~(\ref{eq: dos_ins}), then yields
\begin{equation}
\Delta\mu(T)=\frac{\alpha-\alpha^\prime}{2}T\ln T
+\frac{T}{2}\ln \left [ \frac{C^\prime\Gamma(\alpha^\prime+1)}
{C\Gamma(\alpha+1)}\right ].
\label{eq: deltamu_ins}
\end{equation}

This form for $\Delta\mu(T)$ plays an important role in the thermal transport
of insulators.  Indeed, an examination of the $L_{11}$ and $L_{12}$ 
coefficients in an insulator, with $T\ll E_g/2$, leads immediately to
\begin{equation}
L_{12}=L_{11}\left [ \Delta\mu + \beta\frac{d\Delta\mu}{d\beta}\right ],
\label{eq: l12_ins}
\end{equation}
which follows by taking the derivative of Eq.~(\ref{eq: mu_equality})
with respect to $\beta$ and using the definitions for $L_{11}$ and $L_{12}$
in the low-temperature limit.  If $\Delta\mu(T)=0$, as occurs at half filling,
then we immediately see that $S=0$ (as we also know must hold due to
particle-hole symmetry).  Furthermore, if $\Delta\mu(T)$ depends linearly on
$T$, the thermopower
vanishes linearly in $T$ as well.  $S$ can approach a constant
value if the low-temperature slope diverges [like in the $T\ln T$ dependence
in Eq.~(\ref{eq: deltamu_ins})].  Divergent behavior can occur if
$\Delta\mu(T)$ depends on $T$ with a power-law less than 1
(and greater than 0), but it seems
unlikely that one could find a circumstance where $S$ will diverge like
$1/T$ at low temperature (note that this argument does not hold for
disordered insulators, that have localized states, so that the DOS
does not vanish at the fermi level\cite{villagonzalo}). Hence the qualitative 
temperature
dependence of the thermopower in a correlated insulator can be determined by the
temperature dependence of the chemical potential (at low $T$). Note that the 
above arguments obviously hold for intrinsic semiconductors as well.

If we use the generic insulator DOS in Eq.~(\ref{eq: dos_ins}) and the
approximation that $\tau(\omega)=\tau_0\rho_{int}(\omega)$, then we
find the following behavior for the transport in an insulator
\begin{eqnarray}
\sigma_{dc}&=&\frac{\sigma_0\tau_0}{4}e^{-\beta E_g/2}
\sqrt{\frac{C^\prime T^{\alpha^\prime}\Gamma(\alpha^\prime+1)}
{CT^\alpha\Gamma(\alpha+1)}}
\cr
S&=&\frac{k_B}{|e|}(\alpha-\alpha^\prime)+{\mathcal O}(T)\cr
\kappa_e&=&\frac{\sigma_0\tau_0}{4}e^{-\beta E_g/2}
\sqrt{\frac{C^\prime T^{\alpha^\prime}\Gamma(\alpha^\prime+1)}
{CT^\alpha\Gamma(\alpha+1)}}                
\cr
&\times&\frac{k_B^2}{e^2}
\left [\frac{E_g^2}{2T}-(\alpha^\prime-\alpha)T\right ].
\label{eq: transport_ins}
\end{eqnarray}
The Lorenz number then diverges like ${\mathcal L}=E_g^2/2T^2$
at low temperature.

In a correlated insulator, we must have $\alpha=\alpha^\prime$, so
the thermopower vanishes linearly as $T\rightarrow 0$.  This occurs 
because the interacting density of states develops a gap due to a pole
in the self energy.  Hence the band edge should have the same power law
as that of the noninteracting system, which is determined by the van Hove
singularity, and must be the same for the upper and lower band edge;  the
only place where this analysis might break down is at the critical
coupling strength, where the gap initially 
opens, but we expect that the power-law
dependence of the band edge may differ from the generic case, but it
still should be equal for the upper and lower branches.

Unfortunately the hypercubic lattice in large dimensions does not behave 
like the generic insulator in the strong-coupling limit.  This happens because
the DOS never vanishes for a finite range of 
$\omega$, but rather approaches zero exponentially for all
nonzero $\omega$ (it is suppressed to zero precisely at $\omega=0$
and $T=0$). Hence
we need to analyze the situation for the Gaussian DOS in more detail.

Inside the ``gap region'' of the DOS, the real part of the self energy is
large, because the self energy develops a pole at $\omega=0$ in a
correlated insulator. Similarly, the imaginary part of the self energy
is very small.  Hence, we can approximately determine the Hilbert transformation
in Eq.~(\ref{eq: g_loc}) as
\begin{equation}
G(\omega)\approx \frac{1}{\omega+\mu-{\rm Re}\Sigma(\omega)}-i\pi\rho
[\omega+\mu-{\rm Re}\Sigma(\omega)].
\label{eq: g_loc_large}
\end{equation}
Then we find from Eq.~(\ref{eq: g_0}) that
\begin{eqnarray}
G_0^{-1}(\omega)&\approx&\omega+\mu+i\{{\rm Im}\Sigma(\omega)\cr
&+&\pi[\omega+\mu-{\rm Re}\Sigma(\omega)]^2
\rho[\omega+\mu-{\rm Re}\Sigma(\omega)]\}.
\label{eq: g_0_large}
\end{eqnarray}
Next, by employing Eqs.~(\ref{eq: g_atomic}) and (\ref{eq: g_0}), we finally
determine that
\begin{equation}
{\rm Re}\Sigma(\omega)\approx \frac{w_1U(\omega+\mu)}{\omega+\mu-(1-w_1)U}
\label{eq: re_sigma_large}
\end{equation}
and
\begin{eqnarray}
{\rm Im}\Sigma(\omega)&\approx& -\pi\frac{(\omega+\mu)^2(\omega+\mu-U)^2}
{[\omega+\mu-(1-w_1)U]^2}\rho\left [
\frac{(\omega+\mu)(\omega+\mu-U)}{\omega+\mu-(1-w_1)U}\right ]\cr
&\times& \frac{w_1(1-w_1)U^2}{[\omega+\mu-(1-w_1)U]^2+w_1(1-w_1)U^2}.
\label{eq: im_sigma_large}
\end{eqnarray}
Now the local self energy of the
correlated insulator has a pole at $\omega=0$, hence we learn that
$\mu_0=(1-w_1)U$ for the correlated insulator, and the interacting DOS is
equal to zero at $\omega=0$ (and $T=0$).  
To lowest order, we find that the scattering time then satisfies
\begin{equation}
\tau(\omega)\approx\frac{[\omega+\mu-(1-w_1)U]^4}{(\omega+\mu)^2(\omega+\mu-U)^2
w_1(1-w_1)U^2}
\label{eq: tau_large}
\end{equation}
which is a quartic dependence on frequency in the ``gap region.''  This implies
that for the hypercubic lattice, the scattering time is much bigger than what
one would guess from the approximation $\tau(\omega)=\tau_0\rho_{int}(\omega)$
(which would be exponentially small).

We find that in our calculations, we usually have to input the imaginary
part of the Green's function and of the self energy by hand in the
``gap region'', because the
numerical precision of the computer does not allow them to be determined
accurately with the iterative algorithm (the difficult step is constructing
the self energy from the difference of the inverse of the local Green's
function and the effective medium, where numerical precision is lost).
For accurate calculations, we need to go to higher order in our expansion
than what was illustrated above,
and since the real part of the self energy is determined reliably in
the iterative algorithm, we construct the imaginary parts of the Green's
function and the self energy from the actual converged value of the real part of
the self energy, rather than employing the asymptotic forms discussed above.  
Even with all
of these tricks, it still is a challenge to determine a smooth $\tau(\omega)$
for the correlated insulator.

There is an advantage to studying the Falicov-Kimball model over the 
more popular Hubbard model,\cite{hubbard}
because the Falicov-Kimball model can be tuned to 
have a metal-insulator transition for any value of $w_1$, simply by
choosing $U$ large enough and fixing $\rho_e=1-w_1$
[$\mu=(1-w_1)U$].  The Hubbard model has
a metal-insulator transition only at half filling, where the thermopower
vanishes due to particle-hole symmetry (this can be broken by introducing
multiband Hubbard models\cite{kotliar}).  Hence we can study effects of the 
correlated metal-insulator transition in the Falicov-Kimball model, that are
inaccessible in the single-band
Hubbard model.  Since real materials typically have
complicated bandstructures, one does not expect them to be particle-hole
symmetric except in very special circumstances.  Once again, the Falicov-Kimball
model can be viewed as a more generic metal-insulator transition for
making contact with real materials.  The only disadvantage is that the 
Falicov-Kimball model is a non-fermi-liquid except in ``noninteracting''
limits (where it is a fermi gas).

\section{Numerical Results}

We present results at $\rho_e=1-w_1$
for three different values of $U$: (i) $U=1$ which is a
strongly correlated metal, that has a dip or kink in the interacting DOS at the
fermi level; (ii) $U=1.5$ which undergoes a metal-insulator transition as a 
function of $w_1$; and (iii) $U=2$ which is a correlated insulator with
a sizable ``gap region''.

\begin{figure}[htbf]
\epsfxsize=3.0in
\centerline{\epsffile{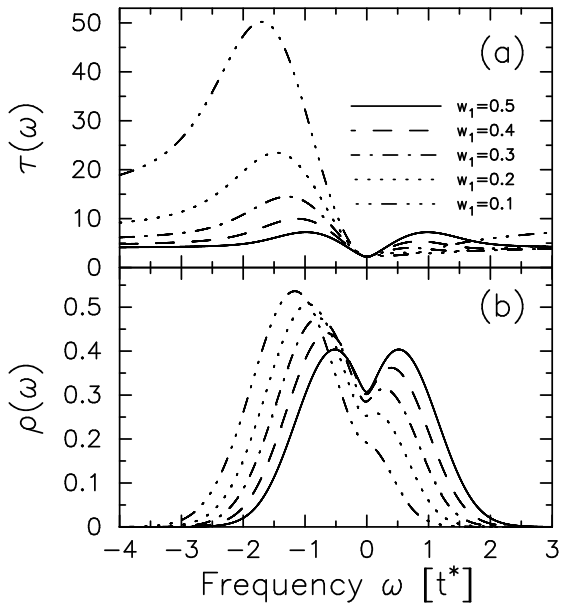}}
\caption{(a) Effective scattering time and (b) interacting DOS for the 
Falicov-Kimball model at $U=1$, $\rho_e=1-w_1$, and $w_1=0.5$ (solid), 
0.4 (dashed), 0.3 (chain-dotted), 0.2 (dotted), and 0.1 (chain-triple-dotted).
Note how both $\tau(\omega)$ and
the DOS develop dips at the fermi level; both functions are symmetric at 
half filling $w_1=0.5$. These plots have fixed the origin at the $T\rightarrow
0$ limit of the chemical potential.
\label{fig: tau_dos_u=1}}
\end{figure}

The effective scattering time and the interacting DOS are plotted in 
Fig.~\ref{fig: tau_dos_u=1}.  We choose $U=1$, $\rho_e=1-w_1$, and vary
$w_1$ (0.5, 0.4, 0.3, 0.2, and 0.1).  Note how the strong correlations
create a dip in both the interacting DOS
and the relaxation time near the fermi level.  As $w_1$ is made smaller,
the scattering time becomes more asymmetric in frequency, which should
yield larger thermopowers. Note that there is residual scattering at the
fermi energy as $\omega\rightarrow 0$, which should produce a finite value
for the DC conductivity at $T=0$ (recall the Falicov-Kimball model is
not a fermi liquid in this regime).

\begin{figure}[htbf]
\epsfxsize=3.0in
\centerline{\epsffile{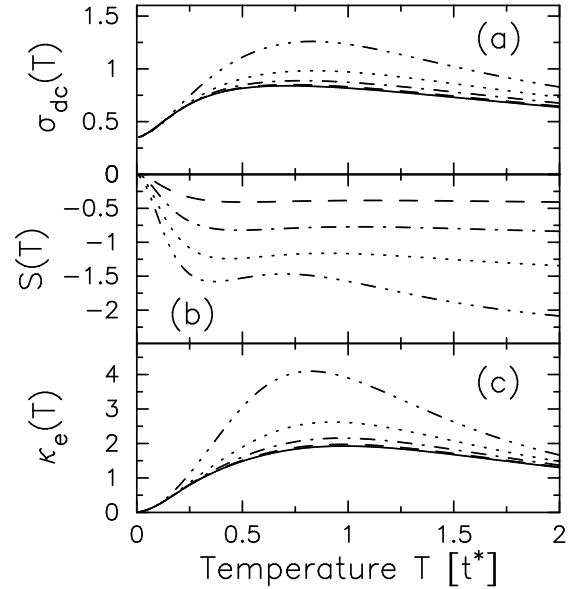}}
\caption{(a) DC conductivity (in units of $\sigma_0$), (b) thermopower
(in units of $k_B/|e|$), and (c) electronic contribution to the thermal
conductivity (in units of $k_B^2\sigma_0/e^2$) for the
Falicov-Kimball model at $U=1$, $\rho_e=1-w_1$, and $w_1=0.5$ (solid),
0.4 (dashed), 0.3 (chain-dotted), 0.2 (dotted), and 0.1 (chain-triple-dotted).
\label{fig: transport_u=1}}
\end{figure} 

We plot the DC conductivity, thermopower, and electronic contribution to the
thermal conductivity in Fig.~\ref{fig: transport_u=1}.  The DC conductivity
is fairly flat as a function of $T$ over a wide temperature range, and is 
enhanced, as $w_1$ is made smaller (and the particle-hole asymmetry is
enhanced).  The thermopower is essentially flat at high temperature, increases
as $w_1$ decreases, and then has a crossover to linear behavior below
$T\approx 0.25t^*$.  The slope of the low-temperature linear behavior 
increases as particle-hole asymmetry increases. The thermal conductivity
is also essentially flat at high temperature, but then has a characteristic
linear behavior that sets in below $T\approx 0.6t^*$.  This behavior is
typical of a metal that has significant scattering.  Note that at low 
temperature, we might expect the Wiedemann-Franz law to hold, while at
high temperature, the Lorenz number must decrease with $T$ (since the
ratio of the conductivities approaches a constant).  The high
temperature decrease of ${\mathcal L}$ with a constant 
(or increasing) thermopower can lead to a large thermoelectric figure of merit.

\begin{figure}[htbf]
\epsfxsize=3.0in
\centerline{\epsffile{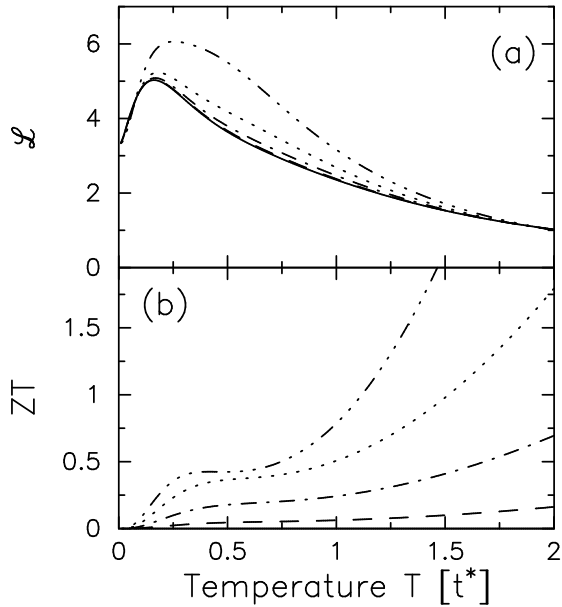}}
\caption{(a) Lorenz number and (b) thermoelectric figure of merit for the
Falicov-Kimball model at $U=1$, $\rho_e=1-w_1$, and $w_1=0.5$ (solid),
0.4 (dashed), 0.3 (chain-dotted), 0.2 (dotted), and 0.1 (chain-triple-dotted).
\label{fig: zt_wf_u=1}}
\end{figure} 

The Lorenz number and thermoelectric figure of merit are plotted in 
Fig.~\ref{fig: zt_wf_u=1}.  Note how the Lorenz number approaches 
$\pi^2/3$ as $T\rightarrow 0$.  This follows whenever $\tau(\omega)\ne 0$
at $\omega=0$, or whenever the interacting DOS at the chemical potential
is nonzero.  The decrease in ${\mathcal L}$ at high $T$ occurs because 
both the DC conductivity and the electronic contribution to the
thermal conductivity become flat at high $T$.  What is interesting is the
moderate temperature peak in ${\mathcal L}$ with the linear decrease toward
the fermi-liquid value as $T\rightarrow 0$.  This occurs because the
system has strong scattering, which produces deviations from the Wiedemann-Franz
law at moderate $T$.  What is unfortunate is that an enhancement of ${\mathcal
L}$ leads to a reduction in $ZT$.  This can be seen clearly in panel (b).
The high-temperature figure of merit is large, and increases in
magnitude as $w_1$ decreases
due to the enhancement of the thermopower (since ${\mathcal L}$ is independent
of $w_1$ at high $T$). But the figure of merit rapidly decreases at low-$T$
and indicates that strongly scattering metals of this type can not be used
for thermoelectric applications at low temperature (but they would work fine
at high $T$ if $\kappa_L$ was small enough, which is usually true).

\begin{figure}[htbf]
\epsfxsize=3.0in
\centerline{\epsffile{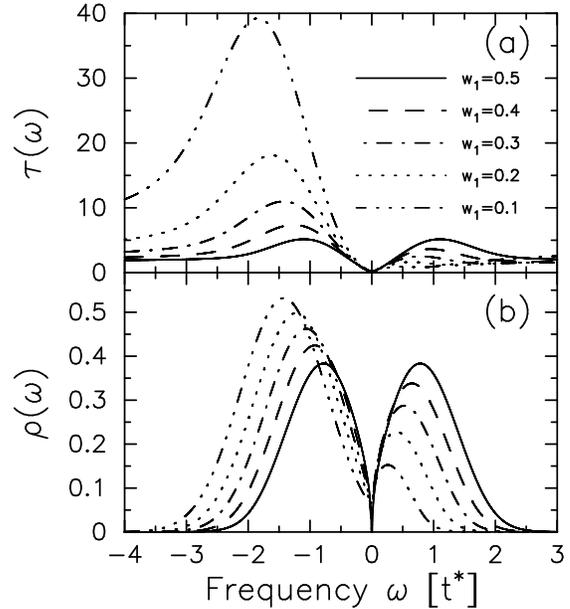}}
\caption{(a) Effective scattering time and (b) interacting DOS for the
Falicov-Kimball model at $U=1.5$, $\rho_e=1-w_1$, and $w_1=0.5$ (solid),
0.4 (dashed), 0.3 (chain-dotted), 0.2 (dotted), and 0.1 (chain-triple-dotted).
Now $\tau(\omega)$ has a quartic dependence on frequency, and the DOS 
becomes exponentially small in the ``gap region''. The gap fills in for 
$w_1< 0.25$ (which is difficult to see in the figure), so we expect the 
transport behavior to be different for small $w_1$.
\label{fig: tau_dos_u=1.5}}
\end{figure} 

Next we show the evolution of the relaxation time and the interacting DOS
as we move into the correlated insulator regime $U=1.5$.  Here the self 
energy develops a pole at $\omega=0$ for $w_1>0.25$, which produces a
narrow region of exponentially small DOS inside a ``gap region'',
but the pole disappears for small enough $w_1$, and the system has strong
scattering, but is still metallic as $T\rightarrow 0$.  This is difficult
to see in the above figure because we are not plotting the gap region
on a logarithmic scale.  We expect that the transport properties may differ
depending on the value of $w_1$ for $U=1.5$, but it should not be too dramatic
because the metallic phase has very strong scattering.

\begin{figure}[htbf]
\epsfxsize=3.0in
\centerline{\epsffile{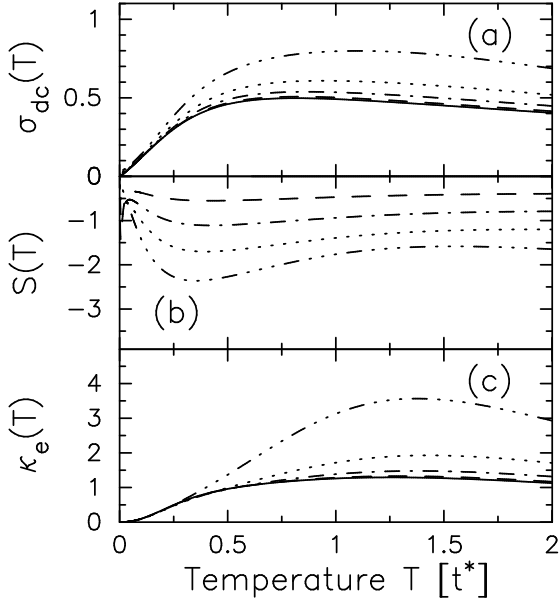}}
\caption{(a) DC conductivity, (b) thermopower, and (c) electronic contribution
to the thermal conductivity for the
Falicov-Kimball model at $U=1.5$, $\rho_e=1-w_1$, and $w_1=0.5$ (solid),
0.4 (dashed), 0.3 (chain-dotted), 0.2 (dotted), and 0.1 (chain-triple-dotted).
Note the very low-temperature peak in the thermopower for $w_1=0.4$ and
$w_1=0.3$.
\label{fig: transport_u=1.5}}
\end{figure} 

The transport coefficients are plotted in Fig.~\ref{fig: transport_u=1.5}.
These look similar to those in Fig.~\ref{fig: transport_u=1} except now the
conductivity vanishes as $T\rightarrow 0$ for the insulating phases and the
thermal conductivity approaches zero faster than linearly for the insulators
as well.  In addition, when $w_1$ is close to half filling, we find a 
low-$T$ peak in the thermopower.  Our calculations indicate that the
thermopower goes to zero as $T\rightarrow 0$, but we cannot rule out the 
possibility of a small nonzero value at $T=0$ (the chemical potential, as
a function of $T$, develops a large negative slope as $T\rightarrow 0$, but
we cannot tell if the slope diverges).

\begin{figure}[htbf]
\epsfxsize=3.0in
\centerline{\epsffile{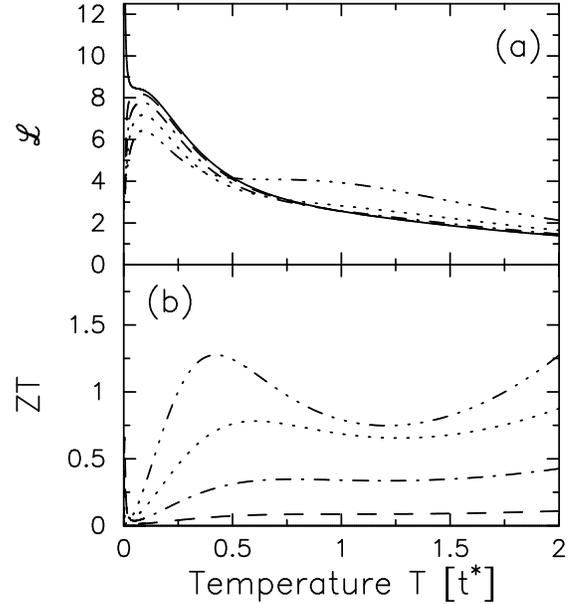}}
\caption{(a) Lorenz number and (b) thermoelectric figure of merit for the
Falicov-Kimball model at $U=1.5$, $\rho_e=1-w_1$, and $w_1=0.5$ (solid),
0.4 (dashed), 0.3 (chain-dotted), 0.2 (dotted), and 0.1 (chain-triple-dotted).  
\label{fig: zt_wf_u=1.5}}
\end{figure} 

\begin{figure}[htbf]
\epsfxsize=3.0in
\centerline{\epsffile{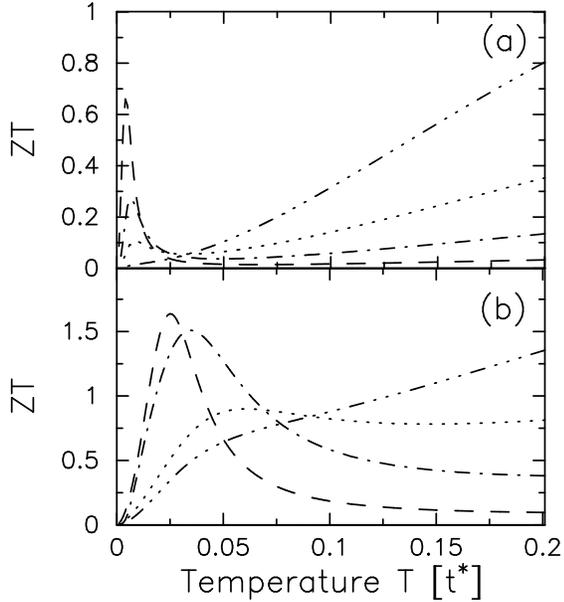}}
\caption{Blow-up of the low-temperature region for the thermoelectric figure
of merit with (a) $U=1.5$ and (b) $U=2$.  The curves correspond to $w_1=0.5$
(solid), 0.4 (dashed), 0.3 (chain-dotted), 0.2 (dotted), and 0.1 
(chain-triple-dotted).  Note how the peak can be pushed to very low $T$
and to reasonably high values when $U$ tuned to lie close to the critical $U$ of
the metal-insulator transition.  The peak generically
moves to higher $T$, increases in magnitude, and broadens as $U$ is increased.
\label{fig: zt_lowT}}
\end{figure}

The plot of the Lorenz number and of the thermoelectric figure of merit 
in Fig.~\ref{fig: zt_wf_u=1.5} show a number of interesting features.
First off, note the very large increase in ${\mathcal L}$ as $T\rightarrow 0$
for the half-filled case versus all other fillings.  This behavior happens
because the half-filled case is qualitatively different from all other cases
at low temperature.  At half filling, the chemical potential is always at
$U/2$ and has no temperature dependence.  Hence the system always has a 
pseudogap-like behavior of the DOS, with it vanishing only at the fermi level.
For all other cases, the chemical potential depends on $T$. Hence, at 
finite-$T$, the DOS at the chemical potential is nonzero, and only vanishes
exactly at $T=0$.  This makes the low-temperature behavior away from half
filling appear like that of a metal, resulting in the Wiedemann-Franz prediction
for ${\mathcal L}$.  Hence there is a low-$T$ downturn to ${\mathcal L}$ 
off of half filling.  Second, at high temperature, ${\mathcal L}$ has doping
dependence now, which is a result of the fact that for $w_1<0.25$ the system
is metallic, while for $w_1>0.25$ it is insulating.  Third, the values of
the thermoelectric figure of merit are enhanced at moderate temperature
for low $w_1$, and there is a low-temperature peak in $ZT$ for $w_1=0.3$ and
0.4, which is hard to see in the figure (it is blown up in 
Fig.~\ref{fig: zt_lowT}). Finally,
the high-temperature behavior of $ZT$ is actually reduced over that of
the correlated metal because ${\mathcal L}$ is somewhat larger and $S$ is 
somewhat smaller here.  $ZT$ is also a flatter function of $T$ here.

\begin{figure}[htbf]
\epsfxsize=3.0in
\centerline{\epsffile{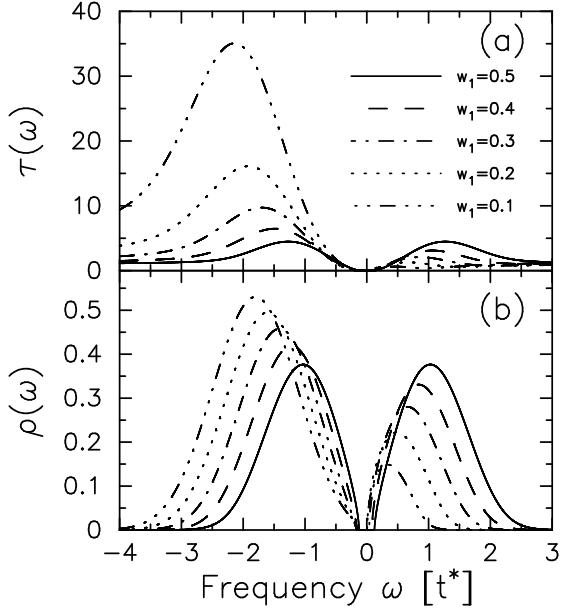}}
\caption{(a) Effective scattering time and (b) interacting DOS for the
Falicov-Kimball model at $U=2$, $\rho_e=1-w_1$, and $w_1=0.5$ (solid),
0.4 (dashed), 0.3 (chain-dotted), 0.2 (dotted), and 0.1 (chain-triple-dotted).
\label{fig: tau_dos_u=2}}
\end{figure}

We plot the relaxation time and interacting DOS for the case $U=2$ in
Fig.~\ref{fig: tau_dos_u=2}.  Now all dopings have well-defined ``gap regions'',
but the size of the ``gap'' is still relatively small.  Note that the
effective relaxation time is quartic for all fillings here.

\begin{figure}[htbf]
\epsfxsize=3.0in
\centerline{\epsffile{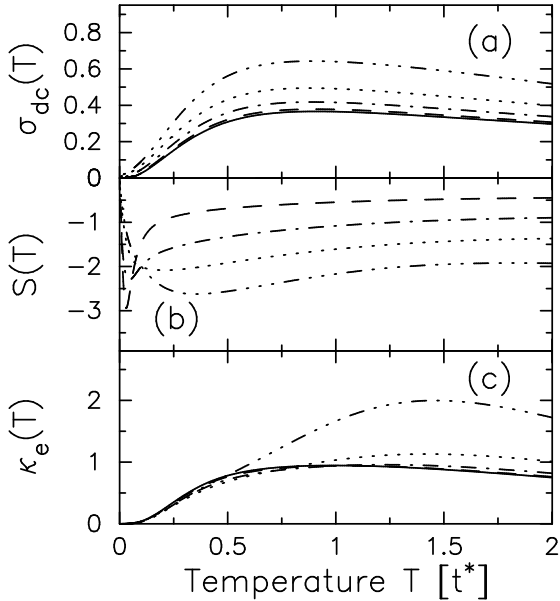}}
\caption{(a) DC conductivity, (b) thermopower, and (c) electronic contribution
to the thermal conductivity for the
Falicov-Kimball model at $U=2$, $\rho_e=1-w_1$, and $w_1=0.5$ (solid),
0.4 (dashed), 0.3 (chain-dotted), 0.2 (dotted), and 0.1 (chain-triple-dotted).
\label{fig: transport_u=2}}
\end{figure}

The transport coefficients are plotted in Fig.~\ref{fig: transport_u=2}
for $U=2$.  Both the electrical and thermal conductivities behave as expected,
with exponentially small values at low $T$ (but the ``gap'' decreases as $w_1$
decreases, so the exponent is doping dependent).  The thermopower still has an
interesting low-temperature peak.  The peak height moves out to larger
$T$ and broadens as $U$ is increased and as $w_1$ is decreased.  We expect this
will have a significant impact on $ZT$.

\begin{figure}[htbf]
\epsfxsize=3.0in
\centerline{\epsffile{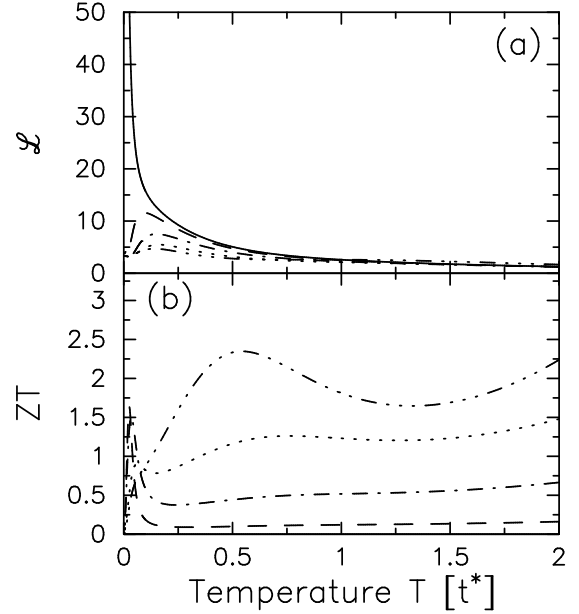}}
\caption{(a) Lorenz number and (b) thermoelectric figure of merit for the
Falicov-Kimball model at $U=2$, $\rho_e=1-w_1$, and $w_1=0.5$ (solid),
0.4 (dashed), 0.3 (chain-dotted), 0.2 (dotted), and 0.1 (chain-triple-dotted).
\label{fig: zt_wf_u=2}}
\end{figure} 

The Lorenz number becomes huge (but does not appear to diverge)
in the correlated insulator at half filling,
but continues to have the low-$T$ peak and return to $\pi^2/3$ as $T\rightarrow
0$ away from half filling.  The thermoelectric figure of merit has a sharp peak
at low temperature, whose width and peak location increase with an increase
of $U$ and a decrease of $w_1$.  Note that the low-$T$ peak is associated with
the sharp drop in ${\mathcal L}$ at low temperature, which seems to occur
only when one has pseudogap behavior and a chemical potential that moves
away from the point where the DOS vanishes at finite $T$.  Nevertheless, the
peak is rather striking and does show that $ZT>1$ is possible in an all
electronic system.  Since the electronic contribution to the thermal
conductivity exponentially decreases at low temperature, while the lattice
contribution decreases as a cubic power law
(when one is well below the Debye energy),
we expect that at very low temperatures
the thermal conductivity will be dominated by a 
lattice contribution, and if that contribution is significantly larger than
the electronic contribution, then the low-$T$ peak in $ZT$ will go away.

We blow up the $ZT$ plots for the low temperature region in 
Fig.~\ref{fig: zt_lowT}.  Note how one can find a sharp peak with an enhanced 
$ZT$ at low temperature when one is close to the metal-insulator 
transition.  As $U$ is increased, the magnitude of the peak increases
and broadens, but it is pushed to higher values of temperature.  We expect
that the lattice contribution to the thermal conductivity may significantly 
reduce this peak, or may even destroy it, but in both cases, the
thermal conductivity at the low-$T$ peak is about $10^{-4}$ the value at high 
$T$ and perhaps the lattice thermal conductivity could be low enough that
it would not significantly interfere with the peak.

\begin{figure}[htbf]
\epsfxsize=3.0in
\centerline{\epsffile{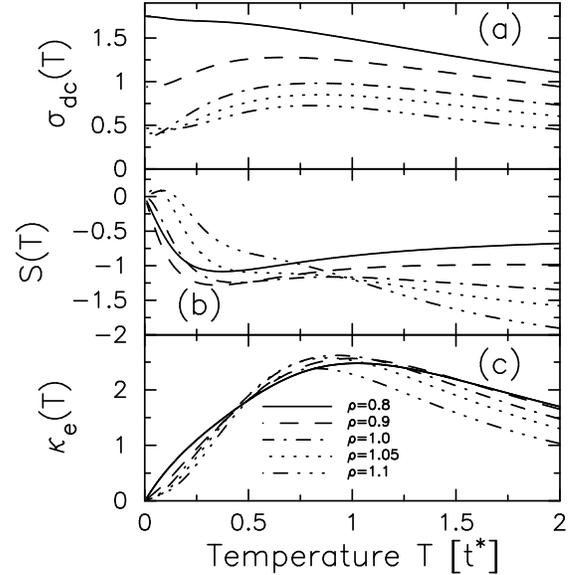}}
\caption{(a) DC conductivity, (b) thermopower, and (c) electronic contribution
to the thermal conductivity for the
Falicov-Kimball model at $U=1$, $w_1=0.2$, and $\rho_e+w_1=0.8$ (solid),
0.9 (dashed), 1.0 (chain-dotted), 1.05 (dotted), and 1.1 (chain-triple-dotted).
\label{fig: transport_u=1_w1=0.2}}
\end{figure}

We conclude our presentation of results by showing what happens when the
system is forced to be metallic, by pushing the electronic chemical
potential away from the ``gap region''.  We start with the strongly correlated
metal at $U=1$ and $w_1=0.2$, but we vary the total concentration to be
0.8 (solid), 0.9 (dashed), 1.0 (chain-dotted), 1.05 (dotted), and 1.1
(chain-triple-dotted).  The relaxation time and the interacting density of
states can be read off of the $w_1=0.2$ curves in Fig.~\ref{fig: tau_dos_u=1},
with the only change an overall shift of the origin for the different electronic
fillings.  The transport coefficients are plotted in 
Fig.~\ref{fig: transport_u=1_w1=0.2}. As expected, the system is least
conductive when the filling is equal to 1 (corresponding to the chemical
potential at the ``kink'' in the interacting DOS).  The thermopower is enhanced
at low temperature as we approach a total filling of one from below
and suppressed at high temperature.  The converse occurs for fillings larger 
than 1---the low-temperature thermopower is reduced and the high-temperature
is enhanced.  This suggests that the thermoelectric properties at high
temperature may be enhanced by increasing the electron filling.

\begin{figure}[htbf]
\epsfxsize=3.0in
\centerline{\epsffile{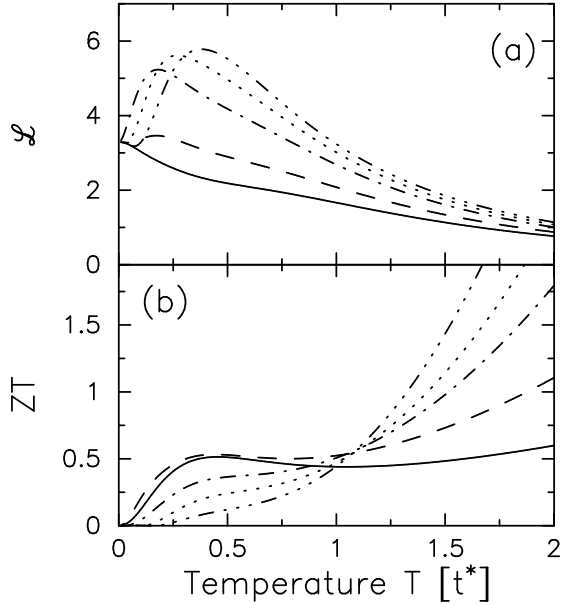}}
\caption{(a) Lorenz number and (b) thermoelectric figure of merit for the
Falicov-Kimball model at $U=1$, $w_1=0.2$, and $\rho_e+w_1=0.8$ (solid),
0.9 (dashed), 1.0 (chain-dotted), 1.05 (dotted), and 1.1 (chain-triple-dotted).
\label{fig: zt_wf_u=1_w1=0.2}}
\end{figure} 

The Lorenz number and thermoelectric figure of merit are plotted in
Fig.~\ref{fig: zt_wf_u=1_w1=0.2}.  Note how the peak in the Lorenz number
only occurs for total fillings larger than about 0.85, and how it gets larger
and moves to higher temperature as the electronic filling increases.  This
has an obvious effect on $ZT$: we find $ZT$ is enhanced at low temperature
for fillings less than 1, and is enhanced at high temperature for fillings
greater than 1.  These effects can be significant.

\begin{figure}[htbf]
\epsfxsize=3.0in
\centerline{\epsffile{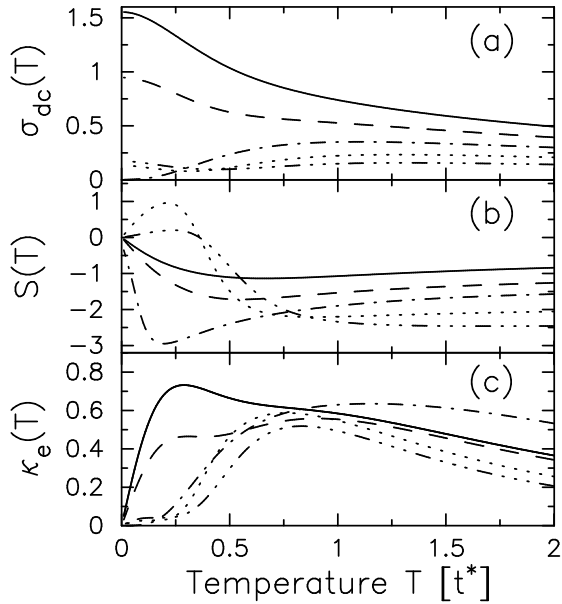}}
\caption{(a) DC conductivity, (b) thermopower, and (c) electronic contribution
to the thermal conductivity for the
Falicov-Kimball model at $U=3$, $w_1=0.2$, and $\rho_e+w_1=0.8$ (solid),
0.9 (dashed), 1.0 (chain-dotted), 1.05 (dotted), and 1.1 (chain-triple-dotted).
\label{fig: transport_u=3_w1=0.2}}
\end{figure}

Finally, we examine a strongly correlated insulator with $U=3$ and $w_1=0.2$.
We don't plot the relaxation time and DOS here, because it is similar to what
is seen for $U=2$, but with a somewhat larger ``gap region''.  The transport
coefficients are plotted in Fig.~\ref{fig: transport_u=3_w1=0.2}.
Here we see a marked difference between the metallic cases, with total
filling not equal to 1 and the insulating case where it equals 1.  In 
particular, the DC conductivity rises as $T$ decreases, as expected for a metal,
but it is exponentially suppressed in the insulator.  The thermal conductivity
is similar, with well-developed linear regimes for the metallic systems (that
are better defined for fillings less than 1), and ``gapped'' behavior for
the insulator.  The thermopower has quite interesting behavior.  For fillings
less than 1, we see a dramatic enhancement in the low-temperature thermopower
as we approach the insulator.  For fillings above 1, we see the thermopower
has a sign change.  This sign change is easy to understand.  When the total
filling lies in the range between 1 and 1.1, the electronic chemical
potential lies in the lower half of the upper Hubbard band at low temperature. 
Hence at low-$T$, the thermopower should appear to be electron-like (positive). 
But as
$T$ is increased, the influence of the gap becomes smaller, and the system
looks overall hole-like because the chemical potential is in the top part
of the overall
``bandstructure''.  The crossover temperature should be on the order
of the size of the ``gap region''.  In addition to this sign change, we
see a large enhancement of the high-temperature thermopower as the filling
increases.  This is also expected because the chemical potential is moving
higher and higher in the band.

\begin{figure}[htbf]
\epsfxsize=3.0in
\centerline{\epsffile{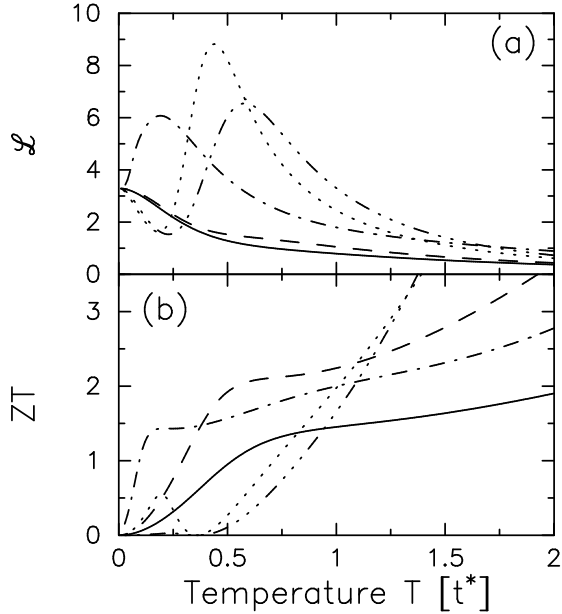}}
\caption{(a) Lorenz number and (b) thermoelectric figure of merit for the
Falicov-Kimball model at $U=3$, $w_1=0.2$, and $\rho_e+w_1=0.8$ (solid),
0.9 (dashed), 1.0 (chain-dotted), 1.05 (dotted), and 1.1 (chain-triple-dotted).
\label{fig: zt_wf_u=3_w1=0.2}}
\end{figure}

We end with the Lorenz number and thermoelectric figure of merit for 
$U=3$ and $w_1=0.2$ in Fig.~\ref{fig: zt_wf_u=3_w1=0.2}.  The behavior here
is also quite interesting.  The Lorenz number is monotonic for total
filling less than one, but then shows a large peak at a filling of 1.  
When the filling increases further, it develops a lower-temperature dip
below a higher-temperature peak.  This low-temperature dip can be advantageous
to thermoelectric properties, but unfortunately the thermopower is too low
(since it passes through zero) in this region to create a large $ZT$.  
Nevertheless, we do see interesting phenomena in the $ZT$ curves as well.
As we approach a total filling of 1, the low-temperature shoulder gets pushed 
closer to $T=0$, and it remains at a fairly high value, above 1. 
When we move past 1, we find the low-$T$ figure of merit is small due to the
proximity to the sign change in $S$, but the high-temperature values are
strongly enhanced.

\section{Conclusions}

We have examined the thermal transport properties of the Falicov-Kimball model
in a binary-alloy picture.  By fixing the ion concentration as a function of
temperature, we force the renormalized energy level of the localized particles
to lie at the chemical potential as $T\rightarrow 0$, which is believed to
be advantageous for thermal transport.  Indeed, we find significant regions
of parameter space with $ZT>1$ at high temperature, and in the correlated
insulators, we also find a small region of parameter space with
$ZT>1$ at low temperature.  Our calculations neglect the lattice contribution
to the thermal conductivity, which should have a limited effect on the high
temperature calculations, but can destroy the low-temperature peaks in $ZT$
if the lattice thermal conductivity is too big.

We showed that generically, in a correlated insulator, one has $S\rightarrow 0$
as $T\rightarrow 0$ and ${\mathcal L}\rightarrow \infty$ as $T\rightarrow 0$.
Our analysis for the thermopower emphasized a relationship between $S$ and
the temperature dependence of the chemical potential, which appears to be
general for systems that display a true gap.
The generic dependences of $S$ and ${\mathcal L}$ at low temperature
imply that the generic thermoelectric figure of merit would be small at
low temperature.  But in the infinite-dimensional hypercubic lattice, the
noninteracting DOS is a Gaussian, which implies that the system really possesses
a pseudogap in the correlated insulator, with the interacting DOS exponentially
small in the gap region.  This has a significant effect on the thermal
transport because the relaxation time is not exponentially small within the
``gap region'', and in particular, it can produce a low-temperature peak to $ZT$
that moves to lower temperature as $w_1$ approaches 0.5 and as $U$ is tuned to 
lie close to the metal-insulator transition.  If we took the hopping 
energy scale $t^*$ to be on the order of 1~eV, then the low-temperature peak
in $ZT$ can easily occur below 50~K [see Fig.~\ref{fig: zt_lowT} where the peak
in panel (a) lies at about $T=0.003$].  The key issue is whether or not the
lattice thermal conductivity would wash out this effect.

One question to ask, then, is can one find a way to introduce an
exponentially small DOS into the gap region of a correlated insulator
(if they do not appear in the bulk system)?  The answer is yes, and it can
be done by creating heterostructures of the correlated material and metals.
The metallic DOS will ``leak'' into the correlated insulator, with a
characteristic length scale, and create small subgap DOS within the system.
Hence we believe the heterostructure idea of Mao and Bedell\cite{mao_bedell} or
Rontani and Sham\cite{rontani_sham} can indeed allow one to get large
peaks in the low-temperature figure of merit (if the exponentially small DOS
also leads to a much larger ``effective''
scattering time).  Since a heterostructure will
also reduce the lattice thermal conductivity, it is possible that the 
low-temperature peak could be realized, but it is likely to require a careful
tuning of the correlation gap, and the thicknesses of the metallic and 
correlated layers of the heterostructure. One also has to be able to maintain
the correlated behavior within the thin layers of the heterostructure.

Another potential way to create states within the gap is simply via a
thermal rearrangement of the DOS within the bulk system as $T$ increases.
In the Falicov-Kimball model, the interacting DOS is $T$-independent, but
in other correlated systems (like the Hubbard or periodic Anderson models)
the correlated DOS does depend on $T$.  It would be interesting to see if
the creation of an exponentially small DOS via thermal activation could allow
for a peak in the low-temperature thermoelectric figure of merit (but this
cannot be studied with the Falicov-Kimball model).

The situation at high temperature is much more promising.  We find that
generically both the electrical and thermal conductivities become flat at
high temperature, so the Lorenz number decreases like $1/T$.  The thermopower
also becomes flat, with a magnitude growing as the system is made more
particle-hole asymmetric.  The net effect is a thermoelectric figure of
merit that grows linearly in $T$, with a potentially large slope.  We found
the figure of merit usually improved at high temperature when the electronic
filling was pushed higher in the band, and that there was no magical need to 
tune the electronic chemical potential to lie in the ``gap region'', rather
the enhancement was generic in correlated systems.

We finally note that our scattering time $\tau(\omega)$ is never close in 
appearance to a delta function in this system.  It can develop large asymmetry
with a large peak lying at one side of the chemical potential, but the
peak width is always broad, and determined by the effective bandwidth
of the conduction electrons.  This is, of course, because we have no
hybridization in this model, which precludes the appearance of a sharp
Abrikosov-Suhl resonance in the DOS and the similar formation of such
a structure in $\tau(\omega)$.  It would be interesting to see how the
situation could change if hybridization was included, but this requires
significantly more sophisticated numerical efforts.

\acknowledgments
This work was supported by the National Science Foundation
under grants DMR-9973225 and DMR-0210717. J.K.F. thanks the hospitality of the
Kavli Institute of Theoretical Physics, where the majority of this work
was completed.  At the KITP, the research was supported in part by the 
National Science Foundation under Grant No. PHY99-07949. V.Z. thanks
the Swiss National Science Foundation grant no. 7KRPJ65554.


\end{document}